\documentclass[aps]{revtex4}
\usepackage{graphicx}
\newcommand{\ini}{\begin{equation}}
\newcommand{\fin}{\end{equation}}
\newcommand{\inia}{\begin{eqnarray}}
\newcommand{\fina}{\end{eqnarray}}

\begin{document}

\draft
\title{\bf De-Sitter and double asymmetric brane worlds}
\author{Rommel Guerrero \footnote{grommel@ucla.edu.ve}, R. Omar Rodriguez \footnote{romar@uicm.ucla.edu.ve} and
 Rafael Torrealba
 \footnote{rtorre@uicm.ucla.edu.ve}}
\address{{\it Unidad de Investigaci\'on en Ciencias Matem\'aticas, Universidad
Centroccidental Lisandro Alvarado, 400 Barquisimeto, Venezuela}}
\begin{abstract}
Asymmetric brane worlds with $dS$
expansion and static double kink topology are obtained from a recently proposed method and their properties are analyzed. These domain walls interpolate between two
spacetimes with different cosmological constants. In the dynamic
case, the vacua correspond to $dS$ and $AdS$ geometry, unlike the static case where they correspond to $AdS$ background. We show that is possible to confine gravity on such branes. In particular, the double brane world host two different walls, so that the gravity is localized on one of them.

\vspace{0.5 cm} PACS numbers: 04.20.-q, 11.27.+d, 04.50.+h
\end{abstract}

\maketitle

\vspace{0.3cm}

\section{Introduction}

The idea that our universe could be a hypersurface embedded in a
higher dimensional spacetime is being considered with raising
interest. During recent years, new models
have been explored
\cite{Horava:1995qa,Horava:1996ma,Lukas:1998yy,Lukas:1998qs,Lukas:1999yn,
Arkani-Hamed:1998rs,Antoniadis:1998ig,Arkani-Hamed:1998nn,Arkani-Hamed:1999hk,
Cremades:2002dh,Kokorelis:2002qi} which require either extra dimensions
with compactification scale close to the limit of modern
measurement of gravity (around millimeters) or noncompact
extra dimensions. In the latter case, the visible universe is confined
into a four-dimensional worldsheet (a $3$-brane) embedded
in a larger spacetime. In particular, the
five dimensional Randall-Sundrum model
\cite{Randall:1999ee,Randall:1999vf} has received much attention
since its birth because of its rich structure, which consists in a thin domain
wall in a relatively simple conceptual framework.

Considering our four-dimensional universe as an infinitely thin
domain wall is an idealization, and it is for this reason, that in
more realistic models the thickness of the brane has been taken into account
\cite{DeWolfe:1999cp,Gremm:1999pj,Csaki:2000fc,Kehagias:2000au,Kehagias:2000dg,Behrndt:2001km,Kobayashi:2001jd,Campos:2001pr}.
The thick domain walls are solutions to Einstein's gravity theory 
interacting with a scalar field, where the scalar field is a
standard topological kink interpolating between the minima of a
potential with spontaneously broken symmetry. We find that
the coupled Einstein-scalar field system has singular solutions \cite{Guerrero:2005xx}. This
kind of solutions to the field equations is obtained from the
usual regular thick domain walls, thus  each regular solution is
associated with an irregular one, just as it happens in classical
sine-Gordon and KdV solitonic theory to accomplish complete integrability of these systems. 
In the present paper, we make simultaneous use of the regular and irregular solutions in order to generate novel brane worlds.  

Due to the non-linearity and instability of the gravitational
interactions, the inclusion of the gravitational evolution into a dynamic
thick wall is a highly non-trivial problem. For this
reason, there are not so many analytic solutions of a dynamic thick domain
wall. In fact, to our knowledge, there exist in the literature only two solutions
\cite{Goetz:1990,Sasakura:2002tq}, with background metric on the
brane given by a de-Sitter ($dS$) expansion, which resembles the
Friedmann-Robertson-Walker metric, typical in a cosmological
framework. In this work,
we obtain another analytic solution of a thick brane with a $dS$
expansion, being $dS$ in one side and $AdS$ in other one along the perpendicular direction to the wall.

In the context of static domain wall spacetime, there are
topological defects richer than the standard topological
kinks. These configurations may be seen as defects that host internal
structure, representing two parallel walls or double domain walls, whose properties have been studied in several papers: in \cite{Campos:1998db}, using models described by a complex scalar field; in \cite{Morris:1995zp,Bazeia:1995ui,Edelstein:1997ej}, in models described by two real scalar fields on flat spacetime; and in \cite{Gregory:2001xu}, in brane scenarios involving two higher dimensions on curved spacetime. On the other hand, the models supported on a single real
scalar field, with a $2$-kink profile have only been
considered in \cite{Melfo:2002wd,Bazeia:2003aw}. In the present
work, we shall deal with very specific models, described
by a single real scalar field which couples with gravity on
a conformal geometry with one extra dimension, where the wall interpolates between different $AdS$ vacua.

The purpose of this paper is to address the issue of localizing gravity in some
new types of domain walls. In the next section, we review
a method to obtain novel brane worlds from known ones developed in \cite{Guerrero:2005xx}, and show
that both branes are not physically equivalent. In sections
III and IV, new domain wall solutions are found and analyzed with both a $dS$ expansion, and
double kink topology. Finally, we
summarize our results on section V.

\section{Approach to generate new solutions}
The nonlinearity of the Einstein's gravity theory
interacting with a scalar field represents a strong difficulty when finding exact solutions to the system. In this sense,
several formulations and methodologies
\cite{Skenderis:1999mm,DeWolfe:1999cp,Guerrero:2002ki} have been
reported, where it is possible to reconsider the problem in a simpler
way. Maintaining this focus, in this section, we will apply a
new formulation based on the linearization of one of the
equations of the system \cite{Guerrero:2005xx}.

Consider the coupled Einstein-scalar field system
\begin{eqnarray}
R_{\mu\nu}-\frac{1}{2}g_{\mu\nu}R=T_{\mu\nu} ,\qquad
T_{\mu\nu}=\nabla_{\mu}\phi\nabla_{\nu}\phi-g_{\mu\nu}\left[
\frac{1} {2}\nabla_{\sigma}\phi\nabla^{\sigma}\phi+V(\phi)\right],
\qquad
\nabla_{\mu}\phi\nabla^{\mu}\phi-\frac{dV(\phi)}{d\phi}=0,\label{acoplamiento}
\end{eqnarray}
where $\mu,\nu=0,...,4$.

We wish to find solutions to this coupled system with a planar parallel symmetry, where the scalar field $\phi$ is a function only of the perpendicular coordinate to the wall $x$. The most
general metric of this kind is
\begin{equation}
ds^{2}=f^{2}(x)\left[-dt^{2}+e^{2\beta t}
 dy^{i}dy^{i}\right]+f^{2}(x)dx^{2},\label{metric}
\end{equation}
where $i=1,2,3$. If ${\beta>0}$, we will have dynamic solutions; while if ${\beta=0}$, we will have static solutions. Following the recently proposed method in \cite{Guerrero:2005xx}, let us $g$ be the inverse function of the metric factor $f$. Then the system (\ref{acoplamiento}) has associated an orthogonal set of solutions $\{g_{1}, g_{2}\}$, for the same scalar field $\phi$, where
\begin{equation}
g_{2}(x)=g_{1}(x)\chi(x),\qquad \chi(x)\equiv\int_{x_{0}}^{x}\frac{1}{g_{1}^{2}(\xi)}\:d{\xi}\label{ec
g2}
\end{equation}
and the general solution
\begin{equation}
g(x)=g_{1}(x)\left[c_{1}+c_{2}\chi(x)\right], \qquad c_{1}, c_{2}\in \Re\label{generalsol}
\end{equation}
depends on two arbitrary constants.

We are interested in the realization of brane worlds on the
geometry defined by $g$. In order to do this, we must guarantee that the
spacetime does not present singularities along the extra coordinate
that prevents the normalization of the zero mode of  the spectra of
gravitons, which is physically unacceptable. The possible
singularities in (\ref{metric})  correspond to the zeros that
(\ref{generalsol}) can take in some values of $x$. Considering $g_{1}$ as a continuos function without zeros, there are two cases where this can happen
\begin{description}
\item[1.-] If $c_{1}=0$ and $c_{2}\neq 0$, then $g$ will have a zero in $x_{0}$.
\item[2.-] If $c_{1}, c_{2}\neq 0$, then $g$ will have a zero in $x_{p}\in x$ when $\chi(x) |_{x_{p}}=-c_{1}/c_{2}.$
\end{description}

Therefore, to avoid divergences in  (\ref{metric}), it is only necessary  to
choose appropriately the constants $c_{1}$ and $c_{2}$, so that
its negative quotient does not belong to the range of
$\chi(x)$, as can be deduced from the second condition.

As a final comment, equation (\ref{generalsol}) gives a
mechanism to obtain new solutions to the coupled system from well
known ones, compatible with the same scalar field but with
different potentials. At this point, it is important to remark
that spacetimes obtained from $g_{1}$ and $g$ with metric tensors
$g_{1\mu\nu}$ and $g_{\mu\nu}$ are physically different, because
it does not exist a diffeomorphism among them. In fact, these
spacetimes are connected by a conformal
transformation given by
\begin{equation}
g_{1\mu\nu}=\Omega^{2}g_{\mu\nu}, \qquad \Omega\equiv c_{1}+c_{2}\chi(x).
\end{equation}
In concordance with \cite{Wald:1984rg}, the conformal
transformations should not be understood as diffeomorphisms, but as
applications that allow to connect two different spacetimes with
identical causal structure. In this sense, we can conclude that
the manifolds with metric tensors $g_{1\mu\nu}$ and $g_{\mu\nu}$
are physically non-equivalent.

\section{A new brane world  with a de Sitter expansion.}
Consider the embedding of a thick $dS$ $3$-brane into a
five-dimensional bulk described by the metric (\ref{metric}), with
a reciprocal metric factor given by
\begin{eqnarray}
g_{1}(x)=\cosh^{\delta}(\beta x/\delta), \qquad 1/2>\delta
>0,\qquad \beta >0.\label{f goetz}
\end{eqnarray}
In Ref.\cite{Goetz:1990,Gass:1999gk}, it has been shown that this spacetime is a solution to the
coupled equations (\ref{acoplamiento}) with
\begin{eqnarray}
\phi=\phi_{0}\arctan\left(\sinh\ \beta x/\delta\right),\qquad
\phi_{0} =\sqrt{3\delta(1-\delta)},\label{campo goetz}
\end{eqnarray}
and
\begin{eqnarray}
V_{1}(\phi)=\frac{1+3\delta}{2\delta}3\beta^{2}\left(\cos
\phi/\phi_{0} \right)^{2(1-\delta)}.\label{potencial goetz}
\end{eqnarray}
The scalar field takes values $\pm\ \phi_0\pi/2$ at
$x\rightarrow\pm\ \infty$, corresponding to two consecutive minima
of the potential with cosmological constant $\Lambda=0$, and
interpolates smoothly between these values; with $\delta$ playing
the role of the wall's thickness. It has been shown that this domain wall geometry localizes gravity on the wall in \cite{Wang:2002pk}
and has a well-defined distributional thin
wall limit in \cite{Guerrero:2002ki}.

Remarkably enough, not so many analytic solutions of a thick
domain wall with $dS$ expansion are known. In fact, an exact
solution on manifold with warped geometry is reported in
\cite{Sasakura:2002tq}; and another one is given by
(\ref{f goetz}, \ref{campo goetz}, \ref{potencial goetz}).
Moreover, this last one is the only dynamic domain wall on a
spacetime with conformal gauge geometry encountered in the
literature so far \cite{Goetz:1990}.

We wish to find another thick domain wall with $dS$ expansion in five-dimensions. From equation (\ref{generalsol}), choosing $c_1=1$, $c_2=\alpha$, with $g_1$ given by (\ref{f
goetz}), we obtain
 \begin{equation}
g=\cosh^{\delta}(\beta x/\delta) + \frac{i\
\alpha\delta}{\beta - 2\beta\delta}\ \cosh^{(1-\delta)}(\beta
x/\delta)\ \sinh^{-1}(\beta x/\delta)\ |\sinh\beta x/\delta|\
{}_2F_1[l, k, n, \cosh^2(\beta x/\delta)],\label{newgeneral}
 \end{equation}
 where ${}_{2}F_{1}$ is the hypergeometric function with $l=1/2-\delta,\ k=l+\delta$, and
 $n=l+1$. This solution represents a three-parameter family.
 Then, for simplicity we will consider the case $\delta=1/4$ without the loss of generality
 \begin{equation}
g=\frac{1}{2\beta}\cosh^{1/4}(4\beta x)\ {\cal H}(x),\qquad
2\beta/\varepsilon
>\alpha > 0,\qquad \beta>0,\label{new g uncuarto}
 \end{equation}
where ${\cal H}(x) \equiv 2\beta-\alpha\ i F[2 i\beta
x,2]$, $\ F$ is the incomplete first order elliptic function and
$1/\alpha\neq{\rm Image}\{\chi(x)\}=\{-\varepsilon/2\beta, +\varepsilon/2\beta\}$ in order to prevent singularities in the metric tensor. Specifically for the case $\delta=1/4$, $\varepsilon = 1.31103\ $. 

This spacetime is a solution to (\ref{acoplamiento}) with
\begin{eqnarray}
\phi=\phi_{0}\arctan\left(\sinh\ 4\beta x\right),\qquad \phi_{0}
=3/4,\label{campo goetz uncuarto}
\end{eqnarray}
and
 \begin{eqnarray}
 V(\phi)&=&\frac{21}{8}\ |\cos\ 4\phi/3|^{3/2}\ {\cal H}(\phi)^{2}-\
     6\ \alpha\  |\cos\ 4\phi/3|^{2} \ {\cal H}(\phi)\
     \tan(4\phi/3)- 6\ \alpha^2\ |\cos\ 4\phi/3|^{1/2}.\label{new potencial uncuarto}
 \end{eqnarray}

\begin{figure}[!htb]
\begin{minipage}[b]{0.3\linewidth}
\includegraphics[width=5.55cm,angle=0]{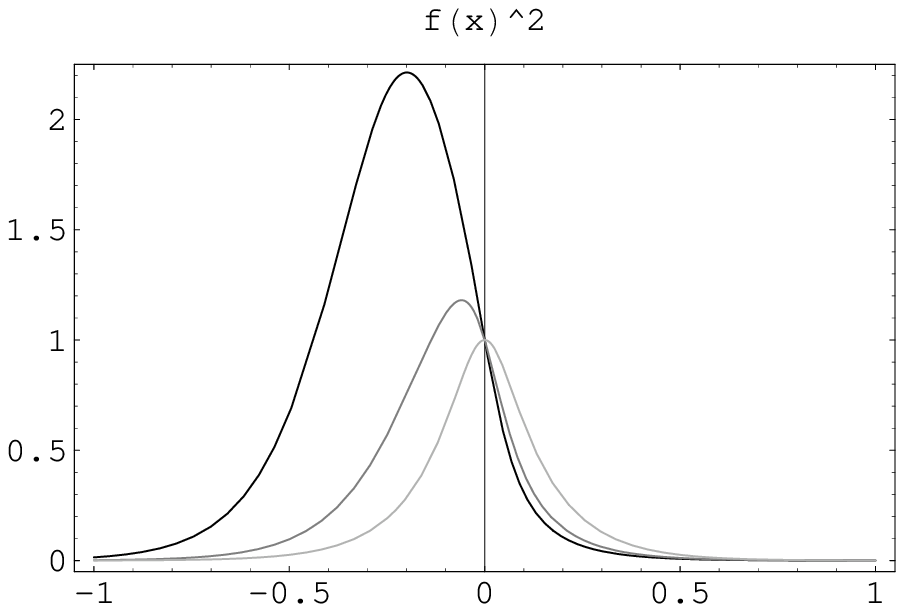}
\end{minipage} \hfill
\begin{minipage}[b]{0.37\linewidth}
\includegraphics[width=5.55cm,angle=0]{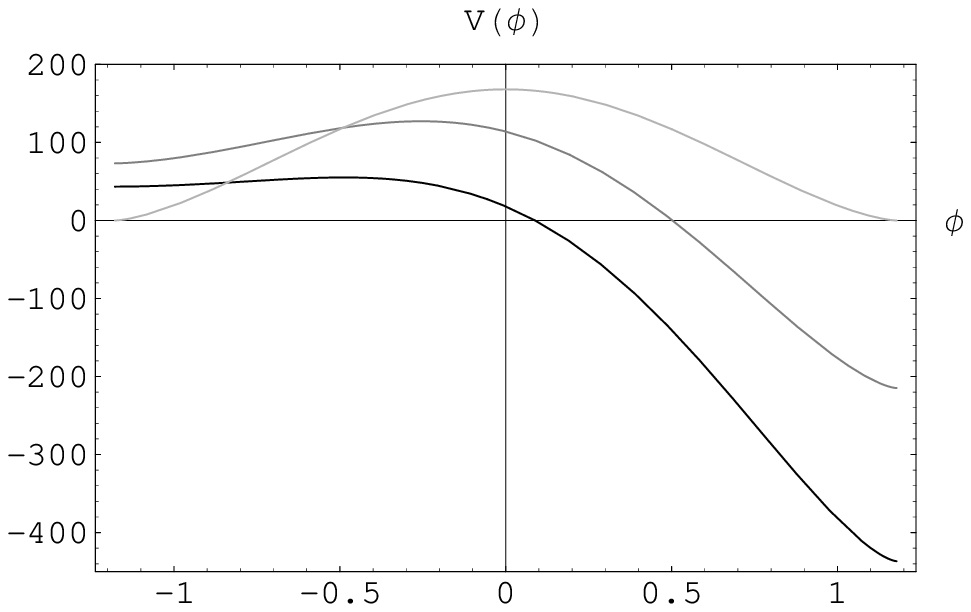}
\end{minipage}\hfill
\begin{minipage}[b]{0.3\linewidth}
\includegraphics[width=5.55cm,angle=0]{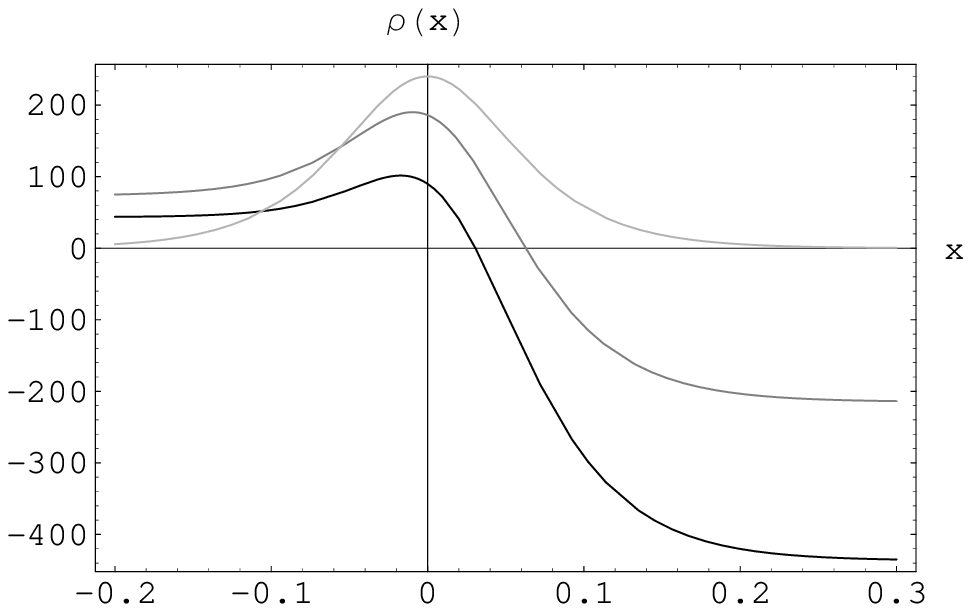}
\end{minipage}\hfill
\caption{Plots of the metric factor (left), scalar potential
(center), and the energy density (right) for $\beta=4$ and
$\alpha\rightarrow0$ and $\alpha=3,5$. Here and in the other figures the thickness of the
lines increases with increasing $\alpha$.}\label{MetricPotential}
\end{figure}

Unlike the domain walls encountered in
\cite{Goetz:1990,Sasakura:2002tq}, for $x\rightarrow +\infty$ this
spacetime is asymptotically $AdS$ with cosmological constant
$-12\alpha\beta(1+\alpha\varepsilon/2\beta)\ $ and for $\
x\rightarrow -\infty\ $ is asymptotically $dS$ with cosmological
constant $+12\alpha\beta(1-\alpha\varepsilon/2\beta)$. The
asymmetry of this solution depends on the $\alpha$ parameter. In fact,
for $\alpha\rightarrow0$ a domain wall is obtained with reflection symmetry
(\ref{f goetz}, \ref{campo goetz}, \ref{potencial goetz}). For
clarity, in Fig.\ref{MetricPotential} we depict the metric factor
$f=1/g$, the potential $V(\phi)$, and the density energy $\rho$
for $\alpha\rightarrow0$ and $\alpha=3,5$.

\begin{figure}[!htb]
\begin{minipage}[b]{0.4\linewidth}
\includegraphics[width=7.5cm,angle=0]{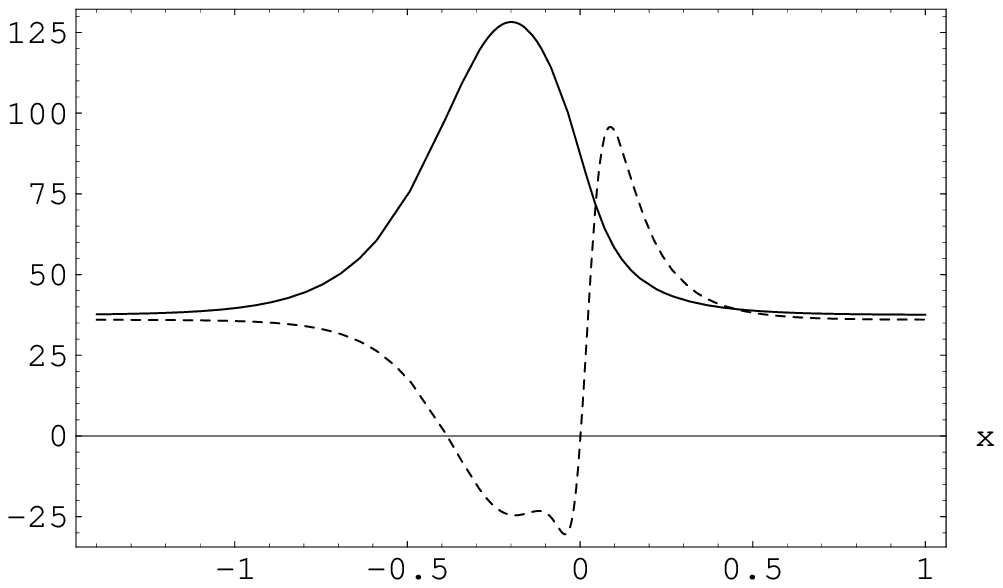}
\end{minipage} \hfill
\begin{minipage}[b]{0.5\linewidth}
\includegraphics[width=7.5cm,angle=0]{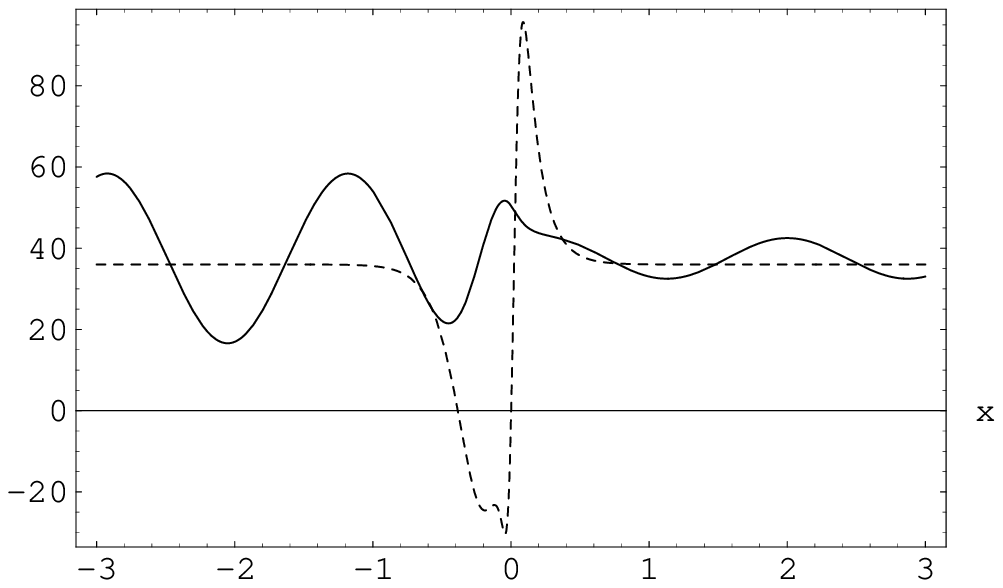}
\end{minipage}
\caption{Plots of the $V_{QM}$ (dashed line), the zero mode (left)
and the massive modes (right) for $\beta=4$, $\alpha=5$ and
$m=7$.}\label{ModesVQM}
 \end{figure}

Next, we shall show that this wall can confine gravity. In
concordance with the straightforward generalization of the
approach used to obtain the general fluctuations around a background
metric presented in \cite{Castillo-Felisola:2004eg}, the equations
which describe linearized gravity in the transverse and traceless
sector are
 \begin{equation}
 -\frac{1}{2}\nabla^{\sigma}\nabla_{\sigma}h_{\mu\nu}+R^{\sigma}{}_{(\mu\nu)}{}^{\varrho}h_{\sigma\varrho}+
R^{\sigma}{}_{(\mu}h_{\nu)\sigma}=\frac{2}{3}h_{\mu\nu}V(\phi),\label{eq
grav lineal}
 \end{equation}
with $h_{\mu\nu}$ the metric perturbations. We write the metric
fluctuations conveniently as
 \begin{equation}
 h_{ab}=e^{i p \cdot y}\ f(x)^{1/2}\
 \psi_{ab}(x),\label{metric perturbations}
 \end{equation}
where $a,b=0,...,3$. From eq. (\ref{eq grav lineal}) we have
 \ini
 (-\partial_x^2+V_{QM})\ \psi_{ab}=m^2\ \psi_{ab},\label{eq.Schrodinger}
 \fin
with $\ -m^{2}=p_{a}p^{a}-2\beta^{2}\ $ the mass in the $dS$ background, and
 \ini
 V_{QM}(x)=\frac{15\beta^2}{\cosh^2(4\beta x){\cal H}(x)^2}\left[\alpha^2\cosh(4\beta x)
 +\frac{\alpha}{2}\cosh^{-1/2}(4\beta x)\sinh(8\beta x){\cal H}(x)-\frac{1}{40}(19-3\cosh 8\beta x){\cal H}(x)^2
 \right],\label{PotencialQM}
 \fin
From eq. (\ref{eq.Schrodinger}) we find that the spectrum of
perturbations consists of a zero mode ($a, b$ indices omitted)
 \ini
 \psi_0=N\left[\frac{1}{2\beta}\cosh^{1/4}(4\beta x)\ {\cal H}(x)\right]^{-3/2},
 \fin
and a set of continuous modes. On the other hand, from
(\ref{PotencialQM}) we can see that $V_{QM}\rightarrow
9\beta^2/4$, as $\ |x|\rightarrow \infty$. Thus the continuous
modes are separated by a mass gap $\ m^2>9\beta^2/4$, as in
\cite{Wang:2002pk}. In Fig. \ref{ModesVQM} we plot the potential
$V_{QM}$, the corresponding zero mode, and the wave
functions for massive modes. We see that the zero mode is bound on the brane, while mass modes move along the bulk, as expected. Moreover, the quantum mechanics potential is such
that, in the transition region $dS-AdS$, there is a potential
barrier in which the low energy massive modes experience a tunnel
effect.

\section{Embedding of a new static thick brane world in an AdS background}

Let us now consider a symmetric thick domain wall spacetime where
the tensor metric is (\ref{metric}) for $\beta=0$, with
 \ini
 g_1(x)=\left[1+(\lambda\:x)^{2s}\right]^{1/2s},\label{metrica
doble}
 \fin
where $\lambda\ $ and $\ s\ $ are real constants with $\ s\ $
an odd integer. This solution was presented and broadly
discussed in \cite{Melfo:2002wd} and represents a two-parameter family of plane
symmetric static double domain wall spacetime, being
asymptotically $AdS_5$ with a cosmological constant $-6\
\lambda^2$. 

The metric with reciprocal metric factor (\ref{metrica doble}) is
a solution to the coupled Einstein-scalar field equations
(\ref{acoplamiento}) with
 \ini
 \phi=\phi_{0}\arctan(\lambda^s\:x^s),\qquad
\phi_{0}=\frac{\sqrt{3(2s-1)}}{s},\label{campo doble}
 \fin
 and
 \ini
 V_1(\phi)=\frac{3}{2}\ \lambda^2\left[\tan(\phi/\phi_{0})\right]^{2(1-1/s)}
\left[\cos(\phi/\phi_{0})\right]^{2(2-1/s)}\left[-1+2s-4\tan^2(\phi/\phi_{0})\right],\label{potencial
doble}
 \fin
where $\phi$ interpolates between the two degenerate minima of
$V(\phi)$, $\ \phi(\pm \infty)=\pm\ \phi_0\pi/2$. Similar solutions are
also considered in \cite{Bazeia:2003aw}.

\begin{figure}[!htb]
\begin{minipage}[b]{0.3\linewidth}
\includegraphics[width=5.55cm,angle=0]{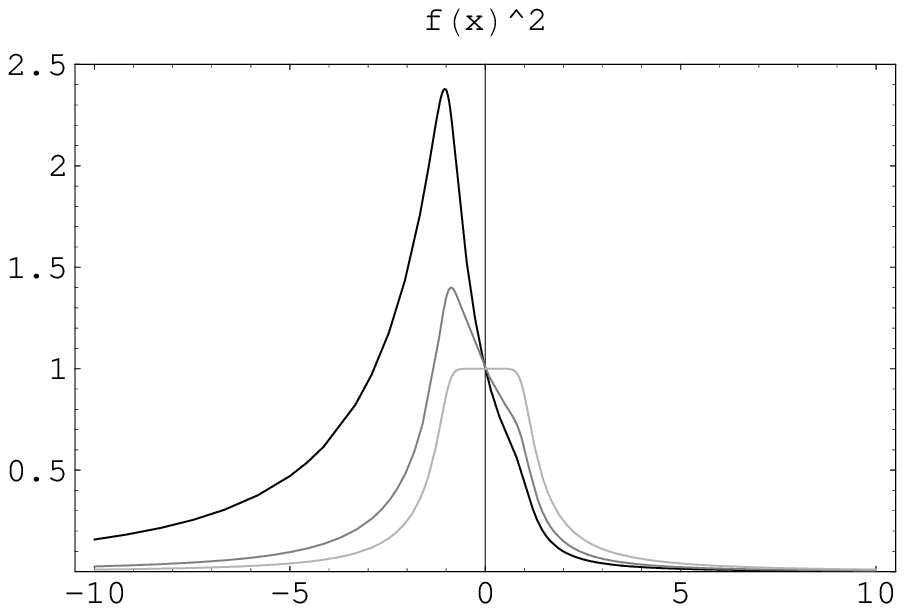}
\end{minipage} \hfill
\begin{minipage}[b]{0.37\linewidth}
\includegraphics[width=5.55cm,angle=0]{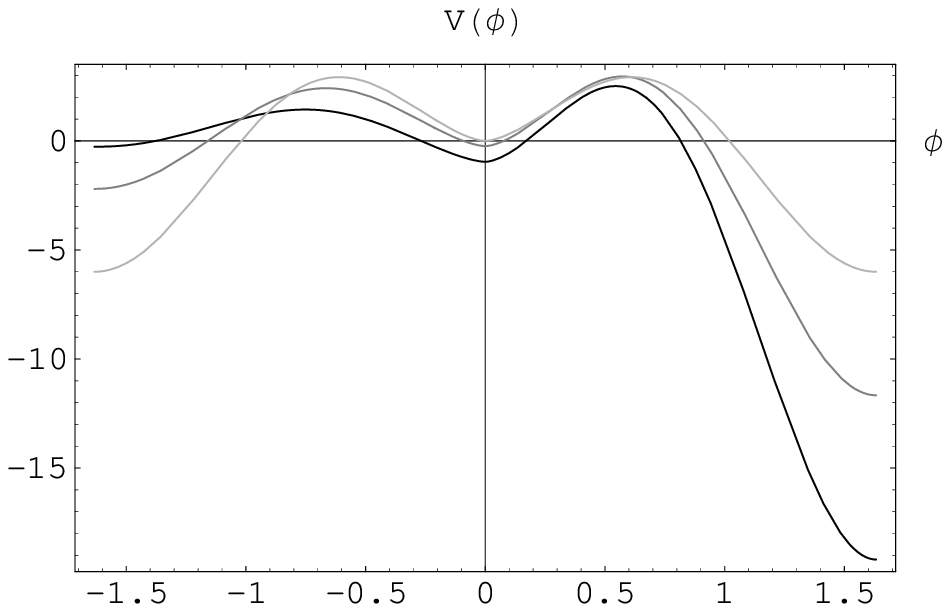}
\end{minipage}\hfill
\begin{minipage}[b]{0.3\linewidth}
\includegraphics[width=5.55cm,angle=0]{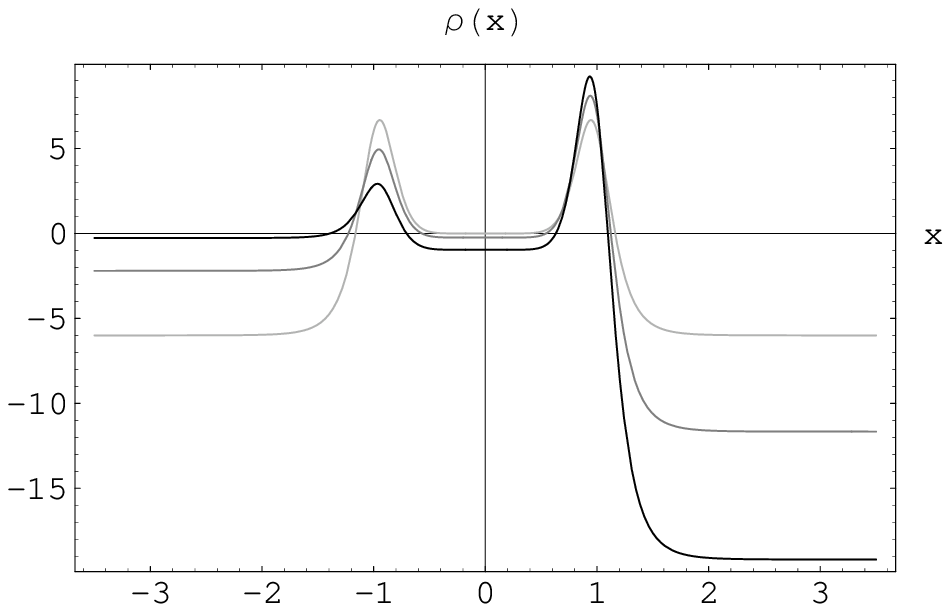}
\end{minipage}\hfill
\caption{Plots of the metric factor (left), scalar potential
(center), and the energy density (right) for $s=5,\ \lambda=1$
and $\alpha\rightarrow0$ and $\alpha=0.2,0.4$.}\label{doubleMetricPotential}
\end{figure}

In concordance with our approach, consider (\ref{generalsol}) with
$g_1$ given by (\ref{metrica doble})
 \ini
 g=\left[1+(\lambda\:x)^{2s}\right]^{1/2s}
 \left(1+\alpha\:x\
 {}_2F_1[l,k,n,-(\lambda\:x)^{2s}]\right),\qquad
 |\lambda|\ \frac{\Gamma(k)}{\Gamma(l)\ \Gamma(n)}>\alpha> 0,\label{g22}
 \fin
where $\ l=1/(2s),\ k=2l\ $ and $\ n=1+l$. This spacetime is also
solution to the coupled system with the scalar field (\ref{campo
doble}) and
 \inia
 V(\phi)&=&-6\alpha^2\cos^{2/s}(\phi/\phi_0)-\frac{3}{4}\sin^2(\phi/\phi_0)\tan^{-2/s}(\phi/\phi_0)
 {\cal K}(\phi)\nonumber \\
 &&\left\{16\alpha\tan^{1/s}(\phi/\phi_0)+\cos^{-2/s}(\phi/\phi_0)
 \left[5-2s-(3+2s)\cos(2\phi/\phi_0)\right]{\cal K}(\phi)\right\}
 \fina
where ${\cal K}(\phi)\equiv\lambda+\alpha\tan^{1/s}(\phi/\phi_0)\
{}_2F_1[l,k,n,-\tan^{2}(\phi/\phi_0)]$. This is a three-parameter
family of plane symmetric static double domain wall spacetime without
reflection symmetry along the direction
perpendicular to the wall. These walls interpolate between $AdS$
asymptotic vacua with different cosmological constant $\ \Lambda_{-}= -6[\ |\lambda|\
\Gamma(k)-\alpha\Gamma(l)\Gamma(n)\ ]^2/\Gamma(k)^{2}\ $ for $\ x<0\
$ and $\ \Lambda_{+}= -6[\ |\lambda|\ \Gamma(k)+\alpha\Gamma(l)\Gamma(n)\
]^2/\Gamma(k)^{2}\ $ for $\ x>0$. For $\ \alpha\rightarrow0\ $ it reduces to
(\ref{metrica doble}, \ref{campo doble}, \ref{potencial doble});
and if, additionally, $\ s=1\ $ the regularized
version of the Randall-Sundrum thin brane is obtained 
\cite{Gremm:1999pj,Guerrero:2002ki}. In
Fig.\ref{doubleMetricPotential} we plot the metric factor, the
potential $\ V(\phi)\ $, and the energy density $\ \rho(x)\ $ for
different values of $\ \alpha$. 

As consequence of the asymmetry of the spacetime (\ref{g22}), we see a difference in the amplitude of the peaks of $\rho(x)$, which indicates that the topological double kink  host two branes whose energy densities differ considerably. Moreover, the lack of symmetry is notably observed in the asymptotically behavior of $\rho(x)$, which reveals the main characteristic of our solution: {\it A double domain wall spacetime interpolating between different $AdS_{5}$ vacua}.

Continuing with the analysis of the system, we would like to explore the thin wall limit of these configurations. Following \cite{Guerrero:2002ki}, we introduce a new parameter $\delta$ by scaling the solution (\ref{g22}), so that the metric is now 
\ini
{}_{\delta} ds^{2}=g(x/\delta)^{-2\delta}\ (-dt^{2}+dy^{i}\ dy^{i})+g(x/\delta)^{-2}\ dx^{2}.\label{dg22}
\fin
Observe that the scaling is performed so that this is still a solution to the Einstein-scalar field equations with
\ini
\phi(x)=\phi_{0}\arctan(\frac{\lambda^{s}x^{s}}{\delta^{s}}),\qquad \phi_{0}=\frac{\sqrt{3\delta(2s-1)}}{s},
\fin
and
\inia
V(\phi)&=&-6\alpha^2\cos^{2/s}(\phi/\phi_0)
 -\frac{3}{4\delta}\sin^2(\phi/\phi_0)\tan^{-2/s}(\phi/\phi_0)
 {\cal K}(\phi)\nonumber\\
 &&\left\{16\alpha\delta\tan^{1/s}(\phi/\phi_0)+\cos^{-2/s}(\phi/\phi_0)
 \left[5-2s-(3+2s)\cos(2\phi/\phi_0)+4(\delta-1)\sin^{2}(\phi/\phi_0)\right]{\cal K}(\phi)\right\}.
 \fina

The metric (\ref{dg22}) is regular in the sense of \cite{Geroch:1987qn}, thus all the curvature tensor fields make sense as distributions. Taking the distributional limit $\ \delta\rightarrow 0\ $ of $G_{t}{}^{t}$ and $G_{x}{}^{x}$ we obtain
\ini
\lim_{\delta\rightarrow 0} G_{t}{}^{t}=-\Lambda_{+}\Theta(x) - \Lambda_{-}\Theta(-x) -3\lambda\frac{(2s-1)}{s}\frac{\Gamma(1-l)^{2}}{\Gamma(2-k)}\left(1+\frac{\alpha^{2}}{\lambda^{2}}\frac{\Gamma(l)^{2}\Gamma(n)^{2}}{\Gamma(k)^{2}}\right)\delta(x); \label{dGtt}
\fin
\ini
\lim_{\delta\rightarrow 0} G_{x}{}^{x}=-\Lambda_{+}\Theta(x) - \Lambda_{-}\Theta(-x),\label{dGxx}
\fin
where $\ \Theta\ $ is the Heviside distribution. This clearly shows that the double configuration, in this limit, corresponds to an infinitely thin domain wall located at $\ x=0\ $, tending asymptotically to $AdS_{5}$ spacetime with different cosmological constant at each side. For the case $\alpha\rightarrow0$, it reduces to the thin wall limit associated to (\ref{metrica doble}) reported in \cite{Melfo:2002wd}.
Remarkably enough, for $\delta\rightarrow 0$, (\ref{dg22}) is not a regular metric in differentiable structure arising from the given chart, and we cannot use the approximation theorems of \cite{Geroch:1987qn} in order to relate the limit of the curvature tensor distributions with the limit of the metric tensor field. Whether or not a metric is regular depends in general on differentiable structure imposed on the underlying manifold. A different chart may exist for which the resulting differentiable structure gives a regular metric, but this is not our concern here.   

Let us return to solution (\ref{g22}) to carry out the analysis of gravitational fluctuations
around this geometry. From eq.(\ref{eq grav lineal}, \ref{metric
perturbations}) with
 \inia
V_{QM}(x)&=&\frac{3\
\lambda\left[1+(\lambda\:x)^{2s}\right]^{-\frac{s+1}{s}}}{4x^2\
{\cal K}(x)^{2}} \left\{5\ \lambda(\alpha
x)^{2}\left[1+(\lambda\:x)^{2s}\right]^{\frac{s-1}{s}}\right.\nonumber\\
&&+(\lambda\:x)^{2s}\left[10\alpha \lambda
x+\lambda\left[1+(\lambda\:x)^{2s}\right]^{-\frac{s-1}{s}}\left[2-4s+5(\lambda
x)^{2s}\right]\right]\nonumber\\
&&+\left.\alpha x(\lambda x)^{2s}\ \left[\lambda+{\cal K}(x)\right]
{}_2F_1[l,k,n,-(\lambda x)^{2s}]\left[10\alpha \lambda
x+\left[1+(\lambda\:x)^{2s}\right]^{-\frac{s-1}{s}}\left[2-4s+5(\lambda
x)^{2s}\right]\right]\right\},
 \fina
we solve for zero modes to get
 \ini
 \psi_0=N\
\lambda^{3/2}\left[1+(\lambda\:x)^{2s}\right]^{-3/(4s)}\
{\cal K}(x)^{-3/2}.
 \fin

\begin{figure}[!htb]
\begin{minipage}[b]{0.4\linewidth}
\includegraphics[width=7.5cm,angle=0]{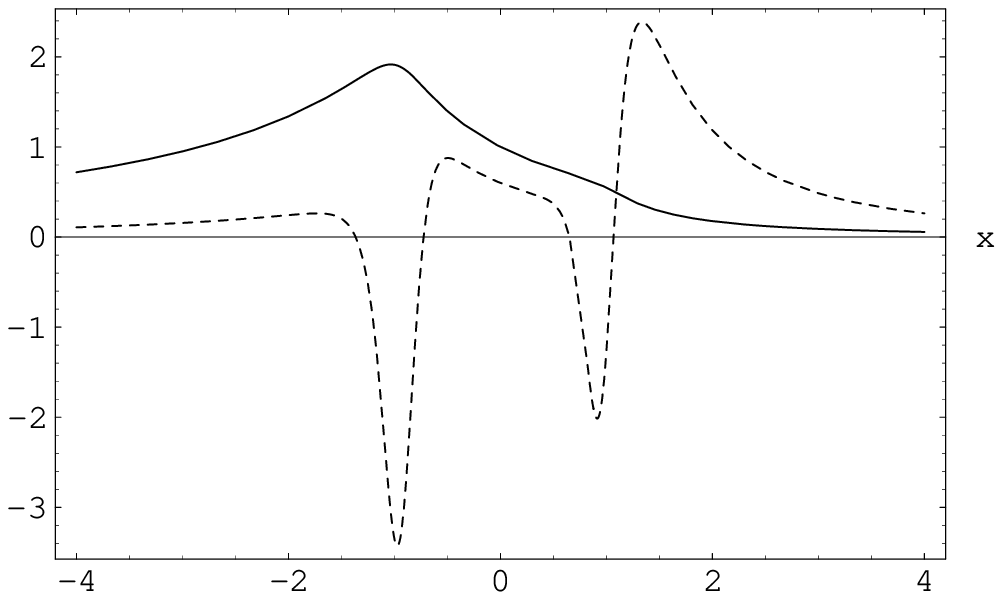}
\end{minipage} \hfill
\begin{minipage}[b]{0.5\linewidth}
\includegraphics[width=7.5cm,angle=0]{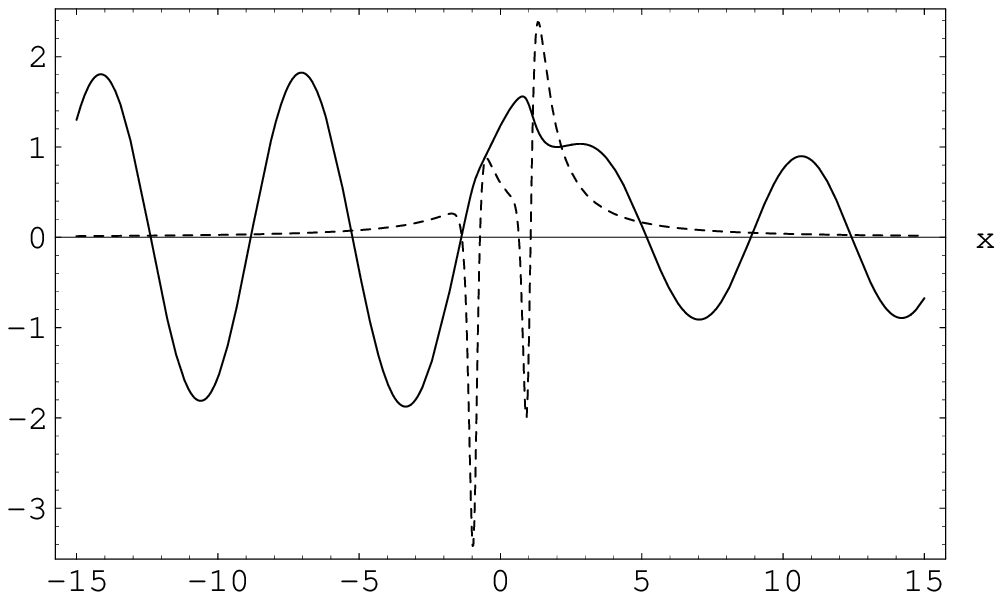}
\end{minipage}
\caption{Plots of the $V_{QM}$ (dashed line), the zero mode (left)
and the massive modes (right)  for $s=5,\ \lambda=1$,
$\alpha=0.4$ and $m=0.9$.}\label{doubleModesVQM}
 \end{figure}

Remarkably, unlike the symmetric double-walls reported in
\cite{Melfo:2002wd,Bazeia:2003aw} where zero mode is essentially
constant between the two interfaces, the massless graviton on our
asymmetric double-walls is strongly localized only on the
interface centered around the lower minimum of the
quantum-mechanic potential, which is in correspondence with the
lower maximum of the energy density. For clarity, in
Fig.\ref{doubleModesVQM} we depict the zero mode and the
potential $V_{QM}$  for the same values of the
parameters. This asymmetric configuration is an exotic example of two parallel walls with different energy densities, where the gravity selects the brane associated to the lower energy conditions as a scenario in order to accomplish our universe. 

In Fig. \ref{doubleModesVQM} we can also see that the wave functions for massive modes move with relative freedom along the extra dimension, where those with the lower energy experience an attenuation due to the presence of the potential barrier localized near the wall with greater energy density. 

Finally, for scaled metric (\ref{dg22}) we find that the linearized equation of motion for tensor fluctuations cannot be rewritten as a Schr\"odringer equation. Moreover, the metric is not regular for $\delta\rightarrow 0$ in the sense of \cite{Geroch:1987qn}. Hence, four-dimensional gravity cannot be reproduced on the thin wall limit and a brane-world in this case is inviable.

\section{Conclusions}
In this paper we find two brane worlds embedded in a spacetime with a non-conventional geometry. 
Both solutions were obtained applying the method developed in \cite{Guerrero:2005xx}, which is supported in
the linearization of one of the equations of the coupling Einstein
scalar field.

In the first case, we studied thick domain walls with $dS$ expansion
in a novel geometry that tends asymptotically to different
cosmological constants, being $dS$ in one side and $AdS$ in the
other one, along the perpendicular direction to the wall. This asymmetry is a consequence of the behavior of the scalar field, which interpolates between two non-degenerate minima of a scalar potential without $Z_{2}$ symmetry. These walls are a generalization
of the first dynamic solution to coupling (\ref{acoplamiento})
reported in \cite{Goetz:1990,Gass:1999gk}. We showed that the
zero mode of the graviton spectrum is localized on the asymmetric brane and there exists a gap between  this state and the massive modes,
which is a generic property of the $dS$ branes.

Finally, in the second case, we considered asymmetric static
double-brane world with two different walls; arising from a scalar potential without $Z_{2}$
symmetry. These branes are embedded  in a $AdS_{5}$ spacetime and in the thin wall limit, the energy density and pressure of these walls correspond to a single infinitely sheet with
different cosmological constants on each side of the wall. The thick
configuration turned out to be a generalization of the regularized
(thick) Randall-Sundrum scenario studied in \cite{Melfo:2002wd}.
We found that the zero mode of the metric fluctuation is
localized on one of the walls, the one corresponding to the conditions of minimum energy; and that the massive modes are not bounded
on any of the walls.  
\section*{Acknowledgments}
We wish to thank A. Melfo and N. Pantoja for fruitful discussion and Susana Zoghbi for her collaboration to complete this paper.
This work was supported by CDCHT-UCLA under project 006-CT-2005.


\end{document}